\documentclass[amssymb,showpacs,showkeys]{revtex4-1}
\usepackage{amsmath}
\usepackage{color}
\usepackage{graphicx}
\begin{document}

\title{Variable Extra Dimensional Spacetime and Solution to Initial Singularity Paradox of Our Universe From Extra Dimensions}

\author{Yong-Chang Huang}
\email[E-mail:]{ychuang@bjut.edu.cn}
\affiliation{Institute of Theoretical Physics, Beijing University of Technology,
 Beijing, 100124, China}

\author{LiuJi Li}
\email[E-mail:]{plilj79@gmail.com}
\affiliation{Department of Physics, University of Naples, Via Cintia, 80126 Naples, Italy}

%\date{\today}

\begin{abstract}
Using a n-dimensional general spacetime and variable extra dimensional spacetime, we achieve its 4-dimensional Einstein equation and  Friedman equations, and discover a general dual relation between the scale factor $a(t)$ of our universe and the scale factor $B(t)$ of extra dimensions. Based on the dual relation equation, predictions of shrinking of extra dimensions and free of singularity problem of our universe are given. Therefore, solution to initial singularity paradox of our universe is achieved. Because the dual relation is general, this Letter discovers that it is just the extra dimensional shrinking contribution that results in our universe's expanding in terms of the dual relation in the bulk space, and actually the deduced dual relation doesn't depend on the 4-dimensional matter concrete Lagrangian, these are key important for a lot of future relative investigations.
\end{abstract}

\maketitle

\section{Introduction}

The idea of extra dimensions plays an important role in theoretical physics.  From the first Kaluza-Klein theory to current M-theory, the concept of extra dimension is widely used to attempt a theory unifying all known fundamental interactions. As well, cosmological aspects of extra dimensional theories have drawn peoples a lot of attentions. The key feature of these theories with extra dimensions is that there are more spatial dimensions compacted to tiny scale, apart from our well-known 4-dimensional spacetime.

Since we have not found any signal of extra dimensions at the current experiments,  size of extra dimensions must be smaller than the extent of current high energy experiments, then it is nature to ask how extra dimensions have been compactified to small scale. There are comprehensive review papers~\cite{OvWe:1997}~\cite{Duff:1994} outlined early works on compactification of extra dimensions.

Usually, cosmological compactification was induced by explicit matter terms, and also, the cosmological constant term is put in these theories by hand. However, Wesson made a suggestion~\cite{Wes:1990} that  5-dimensional Kaluza-Klein equations without sources may be reduced to the 4-dimensional Einstein equations with sources. Next, they investigated a kind of 5-dimensional Kaluza-Klein model whose metric is dependent on the extra coordinate~\cite{Wes:1992}, and its effective 4-dimensional results can be interpreted as  the Scalar-Tensor gravity theory~\cite{FuMa:2003}.

Ref.~\cite{Huang1:2011} gave research on hidden conformal symmetry of extremal Kaluza-Klein
black hole in four dimensions, for applications of string theories with compact extra spacetimes to cosmology, Ref.~\cite{Zhang:2011} investigated DBI potential, DBI inflation action and general Lagrangian relative to phantom, K-essence and quintessence. There were different physical investigatonss about the early universe \cite{KoTu:1990}. In the present work, our aim is to study a more general n-dimensional spacetime by means of a different approach.

Generally, the universe with extra spatial dimensions can be described as a  n-dimensional  spacetime with topology $R^4\times M^{n-4}$, where $R^4$ is  a 4-dimensional spacetime, and $M^{n-4}$ is a ($n-4)$-dimensional space. For the n-dimensional spacetime, we take its general infinitesimal line element square as
\begin{equation}
ds^2=g_{\mu\nu}dx^{\mu}dx^{\nu}+g_{ab} (t,x^a)dx^adx^b,\quad \mu,\nu=0,1,2,3;\quad a,b=1,2,...,n-4,
\end{equation}
where $ds^2(g_{\mu \nu )} =dt^2-a^2(t)\left[\frac{dr^2}{1-kr^2}+r^2d\theta^2 +r^2 sin^2 \theta d\phi \right]$ is corresponding to the Robertson-Walker metrics of our universe, $g_{ab}$ is metric of extra space, in general, it can be functions of $x^{\mu}$ and $x^{a}$, and we consider here the case that it is function of time $t$ and coordinator of extra dimensions $x^{a}$.  We also require $5 \le n \le 11$ following the idea of M-theory in which the total dimensions of supergravity are $11$. For later convenience, we decompose Ricci scalar of the n-dimensional spacetime into a new formula~\cite{Li:2005thesis}
\begin{equation}
\label{ttricsl}
R_{total}  =g^{\mu \nu }R^\alpha _{\mu\alpha \nu } +g^{ab}R^\alpha _{a\alpha b} +g^{\mu \nu }R^c_{\mu c\nu
} +g^{ab}R^c_{acb} =\mathcal{R}+\mathcal{R}'+\mathcal{R}_{cross},
\end{equation}
where
\begin{equation}
\mathcal{R}=g^{\mu \nu }(\Gamma ^\alpha _{\mu \nu ,\alpha } -\Gamma ^\alpha_{\mu \alpha ,\nu } +\Gamma ^\alpha _{\beta \alpha } \Gamma ^\beta _{\mu \nu } -\Gamma ^\alpha _{\beta \nu } \Gamma ^\beta _{\mu \alpha} ),
\end{equation}
\begin{equation}
\mathcal{R}'=g^{ab}(\Gamma ^c_{ab,c} -\Gamma ^c_{ac,b} +\Gamma ^c_{dc} \Gamma^d_{ab} -\Gamma ^c_{db} \Gamma ^d_{ac} ),
\end{equation}
\begin{equation}
\label{rcross}
\mathcal{R}_{cross} =g^{\mu \nu }(-\Gamma ^c_{\mu c,\nu } +\Gamma ^c_{\beta c} \Gamma^\beta _{\mu \nu } -\Gamma ^c_{a\nu } \Gamma^a_{\mu c} ) + g^{ab}(\Gamma^\alpha _{ab,\alpha } +\Gamma^\alpha_{\beta \alpha } \Gamma ^\beta _{ab} -\Gamma ^\alpha _{db} \Gamma^d_{a\alpha } +\Gamma^c_{\beta c} \Gamma^\beta _{ab} -\Gamma^c_{\beta b} \Gamma^\beta _{ac} ).
\end{equation}
This decomposition is similar with the previous work by Straumann~\cite{Straumann:1986} which studied extra dimensions with the structure of a compact Lie group in the framework of principle fibre bundles, but the cross component (\ref{rcross}) is not the same as the one in~\cite{Straumann:1986}  which comes from  Yang-Mills fields.

\section{Cosmology with extra dimensions}

A general action of the n-dimensional spacetime with matter field localized in 4-dimensional spacetime can be written as
\begin{equation}
\label{ndimaction}
\mathcal{I}_n=\int {d^4xd^{n-4}y} \sqrt {-g_{_{total}} }\mathcal{R}_{total}+\int{d^4x} \sqrt{-g} \mathcal{L}_m.
\end{equation}
Using variational principle on the action with respect to $g_{\mu\nu}$, we can easily deduce a 4-dimensional effective Einstein field equation from the n-dimensional spacetime action
\begin{equation}
\label{einsteingm}
R_{\mu \nu } -\frac{1}{2}g_{\mu \nu } \mathcal{R}-g_{\mu \nu }
\frac{\mathcal{R}'+\mathcal{R}_{cross} }{2}=-8\pi G_NT_{\mu \nu } -C_{\mu \nu}.
\end{equation}
where
\begin{equation}
\begin{split}
C_{\mu \nu}=& -\Gamma _{a\nu }^b \Gamma _{\mu b}^a +\Gamma_{\alpha b}^b \Gamma _{\mu \nu }^\alpha +\frac{g^{\alpha \beta}}{2}\Gamma _{\mu b}^b \{\nu ,\alpha \beta \}+\frac{g^{cd}}{2}\Gamma_{\mu b}^b \{\nu ,cd\} -\frac{g^{cd}}{2}\Gamma _{\mu d}^b \{\nu,cb\}\\
&+\frac{g^{cd}}{2}\partial _\mu \{\nu ,cd\}+\frac{g^{cd}}{2}\Gamma_{cd}^\alpha \{\nu ,\alpha \mu \}+\frac{g^{cd}}{2}\Gamma _{\mu\alpha }^\alpha \{\nu ,cd\}-\frac{g^{cd}}{2}\Gamma _{\mu c}^a \{\nu,ad\}\\
&{-\frac{1}{2}\partial _\mu (g^{cd}\{\nu,cd\})} + \frac{1}{2}\partial _\alpha (g^{cd}g_{\mu \nu } \Gamma_{cd}^\alpha ).
\end{split}
\end{equation}

Ricci scalar $\mathcal{R}$ is achieved by timing $g^{\mu\nu}$ on both side of Einstein equation (\ref{einsteingm})
\begin{equation}
\label{ricciscalar}
\mathcal{R}=8\pi G_N T+ C -2(\mathcal{R}'+\mathcal{R}_{cross}) .
\end{equation}
where $C=g^{\mu\nu}C_{\mu\nu}$. Replacing the Ricci scalar in the Einstein equation (\ref{einsteingm}) in terms of (\ref{ricciscalar}), we come to have another formula of Einstein equation.
\begin{equation}
\label{einequ}
R_{\mu \nu} =-8\pi G(T_{\mu \nu } -\frac{1}{2} g_{\mu\nu } T)-(C_{\mu \nu} -\frac{1}{2}g_{\mu \nu } C)-\frac{1 }{2}g_{\mu \nu} (\mathcal{R}'+\mathcal{R}_{cross}).
\end{equation}
There is no new constraint from the contracted Bianchi identities, and the point is discussed in appendix.

Usually, in cosmology, matter in the universe can be considered as perfect fluid whose energy-momentum tensor is $T^{\mu}_{\nu}=\text{diag}(\rho, -P,-P,-P)$, and using Robertson-Walker metric $g_{\mu\nu}$, the effective Friedmann equations of 4-dimensional spacetime are straightforwardly deduced from equation ($\ref{einequ}$)
\begin{equation}
\label{eftfriedequ1}
3\frac{\ddot {a}}{a}+4\pi G_N(\rho +3P) = -\frac{1}{4}\partial _0 g^{ab}\partial
_0 g_{ab} +\frac{1}{4}g^{ab}\partial _0 g_{ab} g^{cd}\partial _0 g_{cd}
-\frac{R'}{2},
\end{equation}
\begin{equation}
\label{eftfriedequ2}
\frac{a \ddot {a}+2\dot {a}^2+2\kappa}{a^2} -4\pi G_N (\rho -P) = -\frac{\dot {a}}{a}g^{ab}\partial _0
g_{ab} -\frac{1}{2}R'\quad .
\end{equation}

\section{Duality between Scale Factors}

On the other side,  recent experimental data~\cite{Riess:1998cb, Perlmutter:1998np} from type Ia supernovae indicate that our universe is accelerating by  dark energy, which can be considered as an effect of the cosmological constant term in Einstein equation.

We now generally generalize the cosmological
constant to a general function of spacetime coordinates. In 4 dimensional theory, the Friedmann equations with the cosmological variable parameter term $\Lambda$ can be obtained from a 4-dimensional action $\mathcal{I}_4=\int d^4x \sqrt{-g}(R+\mathcal{L}_m+\Lambda)$, thus we still have
\begin{equation}
\label{4dfriedequ1}
3\frac{\ddot {a}}{a}+4\pi G_N (\rho +3P)=-\Lambda \quad ,
\end{equation}
\begin{equation}
\label{4dfriedequ2}
\frac{{a \ddot {a}+2 \dot {a}^2+2\kappa}}{a^2} -4\pi G_N (\rho -P)=-\Lambda .
\end{equation}
where $\Lambda(x)$ is a general function of spacetime coordinates. No losing generality, comparing Eq.(13) minuse Eq.(14) with Eq.(11) minuse Eq.(12),  we achieve that right sides of equations (\ref{eftfriedequ1}) and (\ref{eftfriedequ2}) are equal to each other, because the effective Friedmann equations (\ref{eftfriedequ1}) and (\ref{eftfriedequ2}) deduced from higher dimensional spacetime should be the same as the 4-dimensional Friedmann equations  (\ref{4dfriedequ1}) and (\ref{4dfriedequ2}).  It thus yields one equation
\begin{equation}
\label{relequ}
\frac{1}{4}\partial _0 g^{ab}\partial_0 g_{ab} - \frac{1}{4}g^{ab}\partial_0 g_{ab} g^{cd}\partial_0 g_{cd} =\frac{\dot{a}}{a}g^{ab}\partial_0 g_{ab} .
\end{equation}

We consider the condition that extra spatial dimensions are not constant, that is to say, extra dimensions are dependent on time $t$, then the metric of extra spatial dimension can be written as
\begin{equation}
\label{mesd}
g_{ab} =-B^2(t)\tilde {g}_{ab} \quad ,
\end{equation}
where $\tilde {g}_{ab} $ ( a, b = 1,2, {\ldots}, n-4 ) depend only on coordinates of different extra dimensions. Consequently, we have two following equations
\begin{equation}
\label{esdequ1}
g^{ab}\partial_0 g_{ab} =2(n-4)\frac{\dot {B}}{B} ,
\end{equation}
\begin{equation}
\label{esdequ2}
\partial _0 g^{ab}\partial _0 g_{ab} =-4(n-4)\frac{\dot {B}^2}{B^2}  ,
\end{equation}
Substituting equations (\ref{esdequ1}) and (\ref{esdequ2}) into equation (\ref{relequ}),  we thus come up with an equation
\begin{equation}
\label{nrelequ}
(n-4) \frac{\dot{B}}{B} \left((n-3)\frac{\dot{B}}{B}+\frac{1}{2}\frac{\dot{a}}{a}\right)=0
\end{equation}
since $n-4>0$ and $\frac{\dot{B}}{B}\ne 0$ are always valid,  general solution of equation (\ref{nrelequ}) is the same as general solution of following equation
\begin{equation}
\frac{\dot{a}}{a}=-\frac{n-3}{2}\frac{\dot{B}}{B}
\end{equation}
which solution is a relation~\cite{Li:2005thesis}  of scale factors $a(t)$ and $B(t)$
\begin{equation}
\label{relab}
a(t)=\alpha B^{-\frac{n-3}{2}}(t) ,
\end{equation}
where $\alpha$ is an arbitrary constant parameter and we can fix its value later, after we take information of our universe and extra dimensions as input parameters.
Further we get
\begin{equation}
\label{relab}
B(t)={(\frac{\alpha}{a(t)})}^{\frac{2}{n-3}} ,
\end{equation}
Eq.(22) has very important physics meanings, which are given in next section.
\section{Discussion}

The relation (\ref{relab}) exhibits a duality relationship of scale factors between our 3-dimensional space and  extra spatial dimensions. As we know,  physical size of our universe is proportional to scale factor $a(t)$. Since our universe has a long historical expanding, the scale factor $a(t)$ is therefore very large. Then relation (\ref{relab}) indicates that the scale factor of extra dimensions $B(t)$ could be very small for a certain value of $\alpha$, as a consequence, physical size of extra dimensional space is small as well, and it is consistent with the general idea of Kluza-Klein theory and string theory in which extra dimension(s) is(are) compact to very small scale. Furthermore, the relation can be explained as a good example of spontaneous dimensional reduction in  Kaluza-Klein theories, the spontaneous dimensional reduction in any Kaluza-Klein theories always yields a compactified extra-space. The only difference from other approaches is that here the compactification of extra dimensions is not caused by any kind of bosonic matter field coupled to gravity.

However, one may argue that the relation (\ref{relab}) could cause one problem at the beginning of our universe. At the time, because of relation (\ref{relab}) and tiny magnitude of scale factor $a(t)$, extra spatial dimensions could be extremely large. In this case, we would like to give an estimation of the relation equation (\ref{relab}). Updated observational data from Planck collaborator~\cite{Ade:2013zuv} shows that size of our universe is at the degree of $10^{26} m$, and ATLAS group~\cite{Aad:2012fw,ATLAS:2012ky} has released new limitation on the size of extra dimensions, which is smaller than $10^{-16}\ m$.  Taking these two values as input parameters in equation (21), we can get a value $10^{10}$ of constant $\alpha$ when there is only one extra spatial dimension. While, along with increasing of the number of total dimensions $n$, the value of $\alpha$ decreases very fast, and it becomes to $10^{-38}$ when number $n$ is equal to $11$. The big value gap of $\alpha$ does not matter at all, and it shows only that  $\alpha$'s value depends  on the number of extra dimensions $n$.

To see if the condition is reasonable, we choose scale factor $a(t)$ is about $l_{Planck}$ at the very beginning of the universe. If there is only one extra dimension, then its size would be close to $10^{45}\ m$ under the condition $\alpha \sim 10^{10}$. This result is completely surprising, and it is very hard to understand that the size of this large extra dimension is even larger than the size of our current universe, why and how it could shrink to so small size at present from an extreme size at the beginning.  However, when the number of extra dimensions are large, under the conditions $n=11$ and $\alpha=10^{-38}$, the size of extra dimensions  is smaller than $ 0.1 m$, which means that the extra dimensions are small at the beginning of  the universe though its size is extraordinarily big comparing the size of our universe at the time.

As a result, we find that, at the very beginning time, 3 spatial dimensions of our universe are very small, and the extra spatial dimensions are also small. After that, along with expanding of our universe,  the scale factor $B(t)$ of extra dimensions is shrinking. And for this reason, from this relation (\ref{relab}), we can give a safe claim that extra dimensions, in the future, will be harder to find than current, since the size of extra dimensions at the later time will be smaller than its value at present.

In addition, after fixing the value of $\alpha$ in accord with input parameters, we can evaluate the evolution of $B(t)$ with respect to the increasing of $a(t)$ under the conditions of different number $n$.
\begin{figure}
\includegraphics[width=0.5\textwidth]{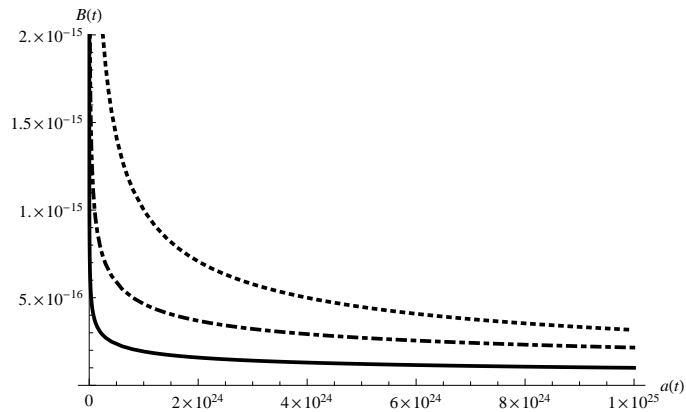}
\caption{Relation of scale factors of our universe and extra spatial dimensions. Dotted line is for n=7, dotted-dashed line is for n=9, solid line is for n=11 }\label{porsf}
\end{figure}
The Fig.\ref{porsf} clearly shows that,  with bigger number $n$, the initial value of $B(t)$ are smaller, and the related $B(t)$ decrease rapidly faster than those $B(t)$ of bigger number $n$. And also, no matter what the number $n$ is at any time, at last, the scale factor $a(t)$ will not go to zero because of the relation (22), i.e., because, in the relation (22), $a(t)$ may infinitely approach to zero but no equal to zero, or the relation (22) cannot keep effect when $a(t)$ is equal to zero and the extra dimensions exist, i.e.,  \emph{the serious mathematical expression, namely the general inner characteristic structure of the bulk spacetime, forbids $a(t)$ to equate zero, but it may infinitely approaches to zero}. Thus there is no the singular property in exact mathematical and physical description. If we accept that there is a minimum limitation,  \emph{Planck length}, in the theory, then size of extra dimensions will go to the limitation if scale factor $a(t)$ increase enough time.

Moreover,  issue of very early singularity paradox of our universe could be avoided under this scenario.  Because it is well known that, at zero scale or below Planck scale, current all physical laws lose effects and actions, which is a real paradox or crisis for all the current well admitted physical theories. For example, in cosmology, in 4 dimensional theories, at the early stage of our universe,  scale factor satisfies $a(t)\sim t^{\frac{2}{3(1+\omega)}}$ and energy density of matter goes to $\rho \sim t ^{-3(1+\omega)}$ from the Friedmann equations and equation of state $P=\omega \rho$. At time $t=0$, scale factor $a(t)$ vanishes and the matter density becomes infinite. This is the singularity problem.  However, the problem will not happen if there exist extra dimensions, because relation (\ref{relab}) forbids scale factor $a(t)$ becoming to zero. It implies that equation of state  $P=\omega \rho$ should be revised at the early universe therefore new physics could emerge.

There are some interesting results from the relation (\ref{relab}), for example, our now 3 spatial dimensions are accelerated expanding because of dark energy or the cosmological constant in current theories, and the relation (\ref{relab}) predicts that extra spatial dimensions are shrinking, but it is not the same as previous works on compactification of extra dimensions which is induced by specific matter terms. In present work, the relation (21) is general and has nothing to do with the 4-dimensional matter Lagrangian, then it is urgent to ask which mechanism leads to this result, it can be seen from the investigation of this Letter what it is just the extra dimensional shrinking mechanism contribution that results in our universe's expanding in terms of dual relation (21) in the bulk space, and that the dual relation is deduced doesn't depend on the 4-dimensional matter concrete Lagrangian, all these are very important for future relative research.

\section{Summary and Conclusions}

 For a n-dimensional general spacetime $R^4\times M^{n-4}$,  we deduce its 4-dimensional Einstein field equation with extra dimensional contribution terms. It leads to an interesting relationship of scale factors between our universe and extra spatial dimensions. And the relation (\ref{relab}) definitely predicts the shrinking of extra dimensions, according to the fact that our universe is expanding. Furthermore, the early singularity paradox can be avoided as a result of dual relation (\ref{relab}), namely, the paradox of the very early singularity of our universe is solved by using the dual relation to forbid the zero singularity of the scale of our universe, because the dual relation must exist when there are extra dimensions. As a further matter, evaluation of the relation (\ref{relab}) shows that value of parameter $\alpha$ strongly depends on the number of extra spatial dimensions, and small value of $n$ might lead to a result that magnitude of extra dimension, at the time as early as \emph{planck time} of our universe, could be ridiculously large. And the dual relation is general, we achieve that it is just the extra dimensional shrinking contribution that leads to our universe's expanding according to the dual relation in the bulk space, and that the dual relation is achieved doesn't depend on the 4-dimensional matter concrete Lagrangian. All the results gotten in this Letter conform to current physics experiments. More theoretical works may be done, which will be written in another paper because of length limit of the Letter.

\appendix*
\section{}
Since it is a n-dimensional manifold, then there are n-dimensional Einstein equations.
 \begin{equation}
 G^{AB}=\kappa_n T^{AB},
 \end{equation}
Its component formulae can be written as
 \begin{equation}\label{4d}
 G^{\mu\nu}=\kappa_n T^{\mu\nu},
 \end{equation}
 \begin{equation}\label{ed}
 G^{ab}=\kappa_n T^{ab}=0,
 \end{equation}
 \begin{equation}\label{4ded}
 G^{\mu b}=\kappa_n T^{\mu b}=0.
 \end{equation}
The last two equations are valid because matter fields are localized in 4-dimensional spacetime.

It is easy to find that   $\nabla_A G^{A b}=\nabla_{\mu} G^{\mu b}+\nabla_a G^{a b}=0$,  because of equations (\ref{ed}) and (\ref{4ded});
 and $\nabla_A G^{A \nu}=\nabla_a G^{a\nu}+\nabla_{\mu} G^{\mu\nu}=0 $, because of equations (\ref{4d}) and (\ref{4ded}). Then the contracted Bianchi identities $\nabla_A G^{AB}=0 $ are always agree with the Einstein equations.

So there is no new constraint on 4-dimensional Einstein equations (10) in the paper.

\begin{acknowledgments}
Authors are grateful for Prof. R. G. Cai for useful discussion. The work is supported by National Natural Science Foundation of China ( grants No. 11275017 and No. 11173028).
\end{acknowledgments}

\end{document}